# The Dark-side of Reionization: Probing Cooling in the Early Universe

Philip Appleton (apple@ipac.caltech.edu)
NASA Herschel Science Center, Caltech

L. Armus (SSC, Caltech), A. Blain (Caltech), F. Boulanger (IAS, Paris), M. Bradford (JPL), V. Bromm (UT Austin), C. Carilli (NRAO), R-R. Chary (SSC-Caltech), E. Egami (Arizona), D. Frayer (NHSC, Caltech), M. Harwit (Cornell), G. Helou (Caltech), M. Lacy (SSC, Caltech), W. Latter (NHSC, Caltech), D. Leisawitz (GSFC), C. Lonsdale (NAASC), A. Ormont (IAP), P. Ogle (SSC, Caltech), M. Ricotti (U. of Maryland), A. Wootten (NRAO)

A whitepaper submitted on 2/14/09 in response to the call from the Astro2010 panel.

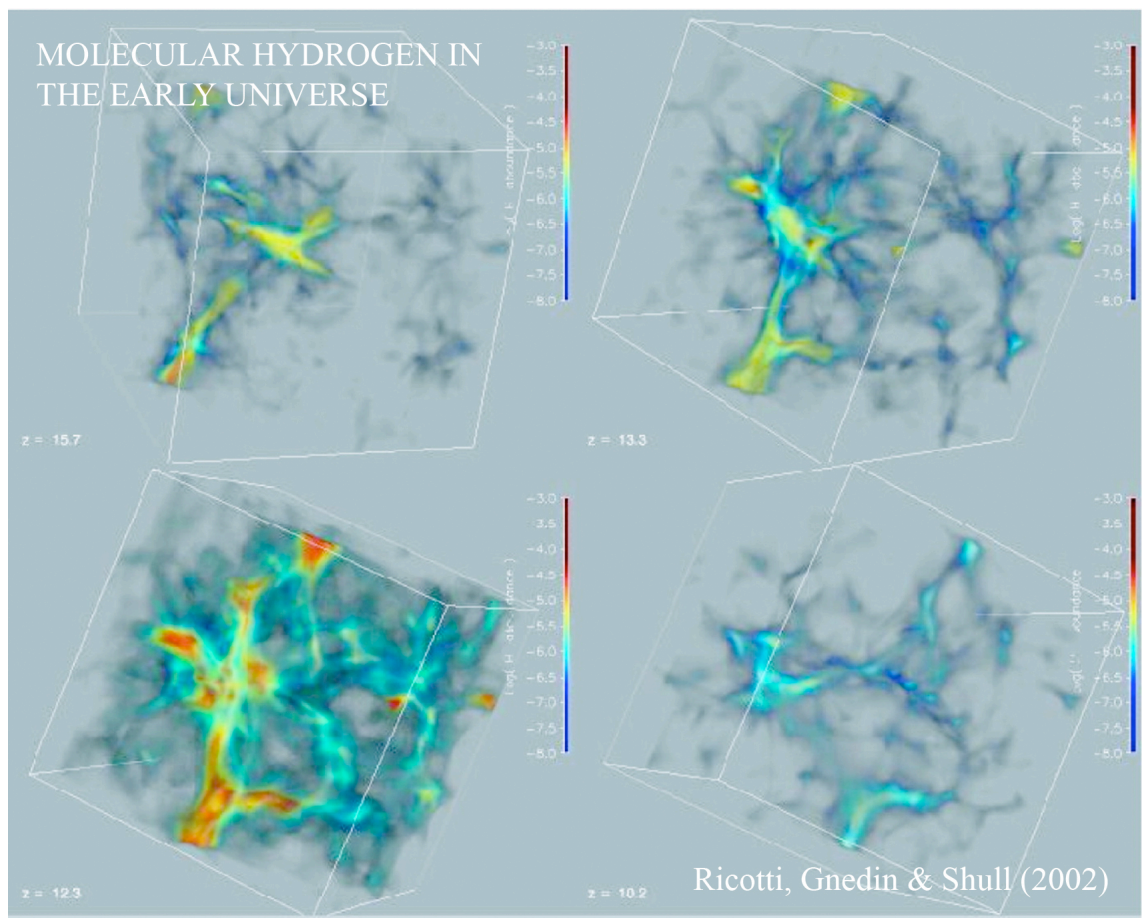

MOLECULAR HYDROGEN IN THE EARLY UNIVERSE

Ricotti, Gnedin & Shull (2002)

**Executive Summary:** Probing the growth of structure from the epoch of hydrogen recombination to the formation of the first stars and galaxies is one of the most important uncharted areas of observational cosmology. During the final stages of this period, the epoch of reionization, the baryonic universe changes from a mainly neutral atomic medium to one which is highly ionized. It marks the transition from the "Dark Ages", to the familiar bright universe of stars and galaxies. We discuss the main molecular and fine-structure metal cooling lines which will be observable in the far-IR and sub-mm emitted by gas-clouds collapsing into growing dark-matter halos during the formation of the first stars and galaxies. These shocked-heated clouds are further stirred by strong feedback effects from supermassive stars, supernovae and hot disks around primitive black-holes. This epoch also marks a transition from a dustless to a dusty medium. Prompt metal-enrichment of the medium by the first massive stars leads to the prediction that fine-structure metal cooling lines will quickly dominate over molecular-line cooling once the clouds cross a critical metallicity. However, new observations of the nearby universe by *Spitzer* have revealed unexpectedly-bright molecular hydrogen lines in galaxies which seem to be powered by turbulent shocks. Since current models of the early universe do not take into account the additional power fed into the $H_2$ through shocks, the $H_2$ lines may be brighter than previously predicted, even where the metallicity is enriched. Although this period of cosmic history is largely unexplored, the faint emission from these first clouds will be observable within the broad-wavelength grasp of the next generation of large aperture space-based far-IR telescopes, and through limited partial-transmission windows in the sub-mm from the ground. With the utilization of gravitational lensing around rich clusters, the signal from high-z protoclouds will be further amplified. The detection of far-IR/sub-mm lines from the "dark side" of reionization will be a perfect complement to the visible emissions from the stars that transformed the Dark-Ages into the age of galaxies. The discovery of the spectroscopic signature of the parent clouds that formed the first stars and galaxies will represent a dramatic leap forwards in our understanding of how our present universe was forged.

**Motivation and Context:** The purpose of this white paper is to draw attention to the very real possibility that important primordial heating and cooling processes occurring during the last stages of the "Dark Ages" can be explored via next generation far-IR and sub-mm spectroscopy. Molecular hydrogen and metal fine-structure cooling lines will provide a missing part of the puzzle of how stars and galaxies emerged from the dark ages. The next few years will see the *James Webb Space Telescope (JWST)* begin to push into the epoch of reionization through visible/near-IR studies of stars and ionized gas surrounding the first galaxies and Active Galactic Nuclei (AGN). At the other wavelength extreme, new radio telescopes like the *LOw Frequency Array (LOFAR)*, and the proposed *Square Kilometer Array (SKA)* will attempt to probe the neutral and ionized hydrogen medium in the dark ages—searching for voids and bubbles caused by the first stars and black-holes[1,2,3]. This paper describes how the far-IR and sub-mm wavebands contain vital clues about the dynamics and thermodynamics of the clouds which led to the formation of the first stars and galaxies. Without this window, we are essential blind to the main cloud cooling processes which inevitably led to cosmic reionization.

Building on the immense success of infrared missions like *Spitzer*, future cooled-aperture IR telescopes will allow us to explore this exciting epoch in cosmic history. Although most theoretical models predict that after metal enrichment by supernovae, metal lines will be dominant, we discuss ways in which the primordial signals from molecular hydrogen might be amplified. This could happen either naturally via the efficient funneling of mechanical energy into enhanced $H_2$-line emission through shocks, or observationally, exploiting gravitational lensing caustics to boost the faint signals. These potential enhancements open-up the possibility that a 3.5 meter-class cooled FIR-telescope might be capable of detecting not just the stronger fine-structure lines of silicon, neon and iron, but also the possibly weaker, primordial $H_2$ lines. A key to the secure identification of these lines in the spectrum will be both the broad wavelength coverage (allowing for multiple-line identifications) and a deep background-limited capability of a new generation of far-IR spectrometers in space. Narrow partial-transmission atmospheric bands (translating into limited redshift coverage) can be explored with the next generation of ground-based sub-mm facilities, some of which could provide maps on sub-kpc scales.

**Tracing Early Chemical Enrichment through Molecular Hydrogen and Atomic Fine Structure Lines:** In the early universe before reionization, baryonic matter collapsed within dark haloes[4] and the stage was set for the formation of the first stars. To form stars, the primordial gas must cool through the formation of molecules, initially $H_2$[5,6] and then below 100K, HD. The main driver for the formation of $H_2$ at z < 100 is via the so-called H-minus route: $H + e^- > H^- + h\nu$ and $H^- + H > H_2 + e^-$ which depends strongly on the free electron density. The tiny free-electron fraction left over after hydrogen recombination at z ~ 1000 can be enhanced in the more massive dark matter halos, leading to rapid $H_2$ cooling and the formation of the first (Population III) stars[7]. Later, at z < 15-20[8], the cosmic background radiation has cooled sufficiently to allow free dust grains (ejected from the first supernovae) to catalyze[9] much more rapid $H_2$ formation on grain surfaces. As more stars form and explode as supernovae, fine-structure lines from metals rapidly dominate over $H_2$ line cooling[10,11]. Spectral line diagnostics in the far-IR and sub-mm will provide a rich and vital new avenue for unlocking the secrets of how the universe was reionized.

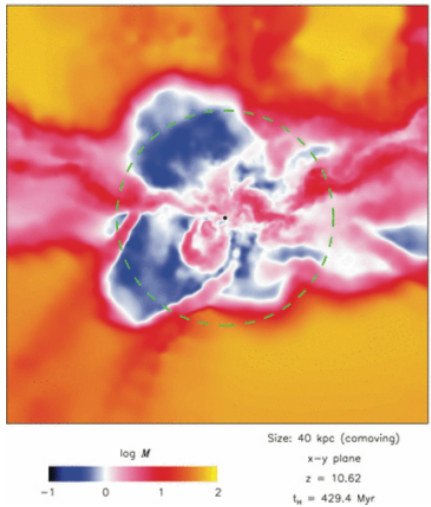

*Fig. 1 Turbulence in the first galaxies[13] The picture shows model conditions within a 1kpc green-dotted circle as in-falling cold streams of gas become highly supersonic (Mach numbers ~10) as they fall into a dark-matter halo. Strong turbulence in a cooling layer behind a shock is seen as a necessary condition to explain the powerful $H_2$ emission from the Stephan's Quintet group and a new population of nearby galaxies which emit very powerful mid-IR $H_2$ emission lines. If present in proto-galaxies, these shock-boosted lines would fall exclusively into the observed far-IR for z < 15.*

**Molecular Hydrogen Cooling :** The build-up of matter into the first galaxies is likely to have been highly turbulent. Massive dark matter (DM) haloes are rare at z > 10, but lower-mass ($10^{8-9}$ $M_{sun}$) haloes are more common. Gas can only collapse into a halo if its temperature is lower than that of the virial temperature of the DM halo. In the absence of heavy elements, $H_2$ will be the primary coolant needed to collapse the gas to form the first stars. The dominant cooling lines are from the lowest-order rotational transition of $H_2$ 0-0 S(0)28μm, 0-0 S(1)17μm, ....0-0 S(5)6.9μm, and will fall exclusively into the far-IR for clouds with z < 15. Ro-vibrational transitions in the rest-frame near-IR (potentially observable with *JWST*) from much hotter $H_2$ are likely to be significantly weaker and will trace a minority of the gas mass. The gas would soon shock and become turbulent as it virializes within the halo[12,13].

**Lessons from the local universe:** Observations of nearby galaxies have revealed some surprises that may be relevant to higher redshifts. Early *Spitzer* pointed-observations of the Stephan's Quintet galaxy group provided evidence of unexpectedly powerful pure rotational $H_2$ cooling from the group[14]. Recent mapping has shown that the $H_2$ emanates from highly turbulent intergalactic filaments caused by a large-scale shock from an intruding galaxy[15]. The luminosity in the $H_2$ lines exceeds the cooling from the fine-structure lines [FeII]26μm and [SiII]34.8μm, as well as the thermal X-ray emission from the shock. The discovery of powerful $H_2$ emission from Stephan's Quintet was quickly followed by even more powerful $H_2$ emission associated with one-in-three local 3CR radio galaxies [16,17]. In 3C326, the integrated emission summed over all the $H_2$ rotational lines amounts to an astonishing 17% of the 18 – 70 μm luminosity of this galaxy. As with Stephan's Quintet, shocks are again implicated: in this case radio jet related. Huge luminosities in the $H_2$ lines have also been reported in several central cluster galaxies at z ~ 0.3[18]. In these cases, the luminosity in the $H_2$ line 0-0 S(1)17μm, is several x $10^{43}$ ergs/s—almost a hundred times the luminosity from Stephan's Quintet. We note that the strength of the $H_2$ emission from the radio galaxies and clusters is such that if they were present at z = 8-10, they would be detectable with future cooled 3-5 meter-class far-IR telescopes. Models suggest that mechanical energy from the shocks is being preferentially channeled into the $H_2$ lines[19], and this is good news for high-z systems where shocks are expected.

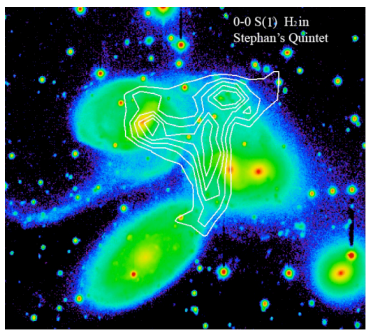
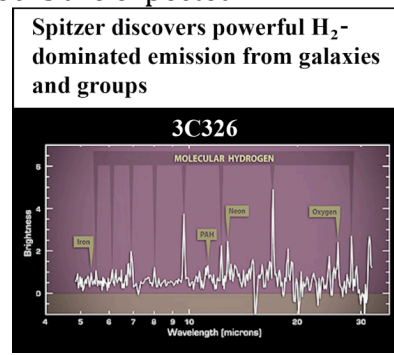

*Fig 2. (Left) Spitzer observations of Stephan's Quintet (SQ) show unexpectedly powerful group-wide 17μm $H_2$ emission (contours) from shocked regions in this compact group[14,15] (Right) The Mid-IR spectrum of 3C326 shows an almost pure $H_2$ spectrum, but is eight times more powerful (8 x $10^{41}$ ergs/s in the $H_2$ lines) than SQ. $H_2$ in 3C326 is likely excited by turbulent shocks caused by a radio jet.*

**Prompt Enrichment and the Primary Metal Coolants:** The conditions in the early universe when metals were absent is believed to favor the formation of very massive stars. This is because the relatively low cooling rates allow large mass-scales to collapse, perhaps biasing these Population III stars towards a top-heavy Initial Mass Function (IMF). However, the injection of metals and dust into the medium from supernovae is likely to have dramatically increased the cooling rate, leading to the formation of predominantly lower-mass stars through fragmentation. The discovery of copious dust emission in z > 6 quasars[20] is testament to the rapid processing of metals in the hosts of these young objects. The explosion of supermassive stars may may provide an explanation, converting as much as 25-30% of the progenitor mass into heavy elements and possibly dust.

The rapid pollution of the primordial gas will lead to a significant new cooling channel dominated by fine-structure line emission of metals[10]. Calculations show high cooling rates due to these fine-structure transitions at redshifts in the range 5 > z > 20 [11]. These lines will be spread throughout the far-IR and sub-mm. Of particular interest are [SiII] and [FeII] lines which can be enhanced in regions where grains are reprocessed into the gas-phase in strong shocks. For example, at z = 6-10, the [SiII]34.8μm line would fall in the far-IR (wavelength range 240-382 μm). [OI]63μm and [CII]158μm lines will be redshifted into the sub-mm/mm range (e. g. 700μm and 1.7 mm at z = 10). Silicate emission from dust grains (rest frame ~ 10μm) from the first SN would be similarly shifted into the far-IR. Therefore to fully comprehend the range of cooling processes, one requires a truly broad wavelength grasp extending over the full range of the far-IR and sub-mm.

We take as our base estimate for the dominant cooling lines theoretical models[11] for a canonical mass clump of $10^8$ $M_{sun}$—the critical scale set when the gas cooling timescale is shorter than the adiabatic timescale of the gas. These values assume emission from a single collapsing clump. The calculations show that after only modest enrichment (~ $10^{-4}$ times solar), the dominant cooling lines are likely to be [NeII]12.7μm, [FeII]26/35.4μm, [SiII]34.8μm (Mid-IR) and [OI]63μm and [CII]158μm (rest far-IR). These lines are predicted to emit in the range ~ $10^{41}$ ergs/s (by coincidence the luminosity of the $H_2$ from Stephan's Quintet), with perhaps more strength from the long-wavelength lines-although this depends on grain processing. Fig. 3 shows how some of the line luminosities translate into potentially observable line fluxes for clouds at 6 < z < 10. In the absence of shock boosting, the predictions for the 28 and 17 μm $H_2$ lines are much lower than the metals, being a factor of more than 100 times fainter once the metals are formed ~$10^{39}$ ergs/s. Higher fluxes, exceeding $10^{41}$ ergs/s are only possible if extreme assumptions are made about the star formation rate in these clouds[21,22] or if the kind of shock-boosting seen in Stephan's Quintet can be applied to these systems. The predictions are for a single clump.

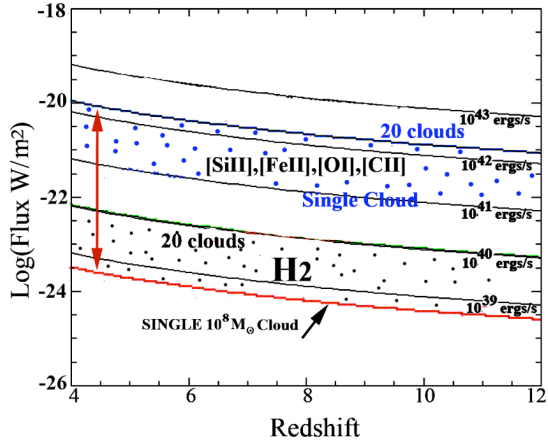

*Figure 3. $H_2$ (T=200K) and Fine-Structure Line flux predictions for single[11] and multiple cloud condensation of $10^8 M_{sun}$. The lower band shows the fainter predictions for $H_2$ after metal pollution and the upper band shows the stronger emission expected from fine structure lines. These lines will be spread throughout the far-IR and sub-mm for high redshift emission. We assumed a $\Lambda$CDM cosmology ($\Omega_L$=0.7, $\Omega_m$=0.3, $H_0$=70).*

In reality, simulations show a more complicated picture in which these clumps form "beads on a string" along cosmic filaments[23,0]. Complicated heating and cooling effects[24,25,26] and $H_2$ reformation can lead to highly inhomogeneous molecular distributions, and sites for the first stars. Observationally (assuming a $\lambda$200µm telescope beam of ∼ 12 arcsecs for D = 3.5 m) we might expect to observe some lines of sight containing many $10^8$ $M_{sun}$ clumps and others with few. Fig. 3 shows the line flux predictions for single and multiple cloud condensations. The predictions for $H_2$ do not include possible shock excitation of the $H_2$.

**Prospects for Far-IR Detection From Space:** Building on the remarkable success of *Spitzer*, future generations of cooled far-IR telescopes equipped with new generation detectors will provide a remarkable opportunity for astronomers to explore molecular hydrogen and fine-structure metal lines in the epoch of galaxy formation. Line widths of ∼50-200 kms$^{-1}$ might be reasonable given the halo masses expected, but more theoretical work will be needed to predict detailed redshift distributions. We will require i) the ability to obtain spectra with spectral resolution of R∼ 1000 over a large wavelength range (100-500µm) so that multiple lines can be identified. This is especially important given the wide range of possible redshifts that these clouds might inhabit along a particular line of sight, and ii) the ability to push as close to the theoretical background limit as possible by exploiting both large collection areas (to beat foreground galaxy confusion), and cold mirror technologies. Warm-mirror facilities like *Herschel* will make great strides in opening up the far-IR/sub-mm window, but fall far short of the depth needed to detect the primordial lines shown in Fig.3. We therefore look to both near-term (the 3.5 meter proposed Japanese Space Infrared Telescope for Cosmology and Astrophysics (SPICA) mission), and long-term future larger aperture platforms like *Single Aperture Far-IR Observatory (SAFIR)*, as well as limited wavelength coverage from the ground from the *Atacama Large Millimeter Array (ALMA)* and the *Cornell Caltech Atacama Telescope (CCAT)*.

An instrument like the *Background-Limited Infrared-Submm Spectrometer (BLISS)* medium-resolution (R~700) spectrometer (www.submm.caltech.edu/BLISS/bliss_poster_small.pdf) attached to SPICA could reach the lines fluxes shown in Fig. 3 for the primordial clouds. The case for the fine structure lines seems the most encouraging. Achievable line-sensitivities for SPICA-BLISS (5$\sigma$-1hr = 1 x 10$^{-20}$ W/m$^2$), or SAFIR, a possible 6-8-m class cooled telescope in space and assuming reasonable advances in detectors (5$\sigma$-1hr = 5-10 x 10$^{-22}$ W/m$^2$) show potential for detection of the fine-structure lines. As Fig 3 shows, H$_2$ emission is expected to be more difficult to detect unless the kind of enhanced H$_2$ emission seen in the local universe (see section 3.3) can be translated to high-z. The [OI] and [CII] lines will be tackled by ground-based telescopes. To put things in perspective, SPICA/BLISS would reach sensitivity levels ~300 times deeper than the *Herschel/PACS* instrument, 1000 times deeper than an equivalent instrument on the *Statospheric Observatory for Infrared Astronomy* (*SOFIA)*, and 10 x deeper than CCAT over compatible wavelength ranges.

**Gravitational Lens Boosting**: Although it is conceivable that the metal lines might be detected directly by SPICA/BLISS as Fig. 3 shows, the H$_2$ lines are likely to be challenging unless a means can be found to provide further signal boosting. If the clouds are point-like, as seems suggested from simulations of these bead-like condensations, it may be possible to boost the signal around strong gravitational-lensing caustics. This has two benefits: 1) It could provide an order of magnitude boosting or more if cloud-clumps were to fall along a caustic, and 2) it reduces the confusion of foreground star forming galaxies along sight-lines to the clumps: an important consideration in the 100-300μm regime where the lines are likely to fall. Predictions for the collapse of ~10$^8$ M$_{sun}$ halos suggest that shocks with strong H$_2$ emission will form within the virial radius at a few hundred pc[13]. Thus the brightest emission will occur on relatively small physical scales << 1kpc. Such objects are prime candidates for magnification by galaxy cluster lenses with telescope beam sizes of 5-12 arcsecs. There are of order 50 clusters known with redshifts and masses suitable for use as lenses. Calculating the projected area along lensing caustics we find that more than a factor of 10 amplification is expected along critical lines, corresponding to an effective survey volume of ~1000Mpc$^3$, for 6 < z < 10. This provides a clear strategy for targeting background protoclouds.

Objects found in future deep surveys of strong lensing clusters by *JWST* may even provide specific targets for a *SPICA/BLISS* type instrument. Given the number density[27] of dark haloes ~10$^9$ M$_{sun}$ at z~8 is about 1/Mpc$^3$, and that 1% of the clumps in such halos might be in the (~ 10$^7$ yr) collapse/starburst phase, we can expect at least 10 objects of the desired size and scale would be amplified in such a lensing survey. This demonstrates that using lensing surveys to boost the protocloud signal will lead to a positive benefit for telescopes in the 3.5-m SPICA class.

**Ground-based Detection in the Sub-mm:** Several partial-transmission bands exist in the sub-mm that could be explored from the ground through restricted atmospheric windows. The 315-380 micron band offers one such window available at high altitude sites such as those of CCAT and ALMA. CCAT's wide field of view is complemented by ALMA's high spatial resolution. Observations may be carried out during about 25% of the time with, for example, ALMA sensitivity of 5$\sigma$-1hr of 10 mJy (for a 300 km/s linewidth this corresponds to 8.5 x 10$^{-20}$ W/m$^2$ for comparison with Fig. 3). Under the very best conditions, one might improve this by a factor of three. Under these same exceptional conditions, ALMA could observe in a 200-240μm band, with an expected sensitivity of 5σ-1hr of 10 mJy (or 1.4 x $10^{-19}$ W/$m^2$ for linewidth = 300 km/s). Considerably improved sensitivities are obtained with ALMA in the 2mm band where highly redshifted [CII]157um emission might be expected. At these long wavelengths, the Milky-way galaxy could be detected at z = 10 in a 4hr transit. Similar sensitivities would be expected for [OI]63μm line. ALMA is well suited for high spatial-resolution studies of high-z long-wavelength galaxy transitions shifted into the sub-mm/mm.

**Summary:** Far-IR spectroscopy covering λ100-500μm from space, and narrow partial-transmission atmospheric bands available from the ground, opens up the possibility of probing the molecular hydrogen and metal fine-structure lines from primordial clouds from which the first stars and galaxies formed at 6 < z < 15. Building on *Spitzer* observations of unexpectedly powerful $H_2$ emission from shocks, we argue that next-generation far-IR space telescopes may open a new window into the main cloud cooling processes and feedback effects which characterized this vital, but unexplored epoch. Without this window, we are essential blind to the dominant cloud cooling which inevitably led to star formation and cosmic reionization.